\documentclass[12pt,a4paper]{article}
\begin{document}
\large
\author{Yu.A.Mamedov, H.I.Ahmadov \footnote{E-mail: hikmatahmadov@yahoo.com} \\
Chair of Mathematical Physics, Faculty of Applied Mathematics\\
and Cybernetics, Baku State University Z.Khalilov st.23, 370148\\
Baku, Azerbaijan}
\title{Almost regularity conditions of spectral problems \\
for a second order equation}
\date{}
\maketitle
\begin{abstract}
In this paper the asymptotic distributions are exactly solved for linearly
independent solutions considering problem of the second order and for the
coefficients of asymptotic destribution the recurent formulas are obtained.
Further, using obtained recurent formulas the neccessary and sufficient
conditions almost regularity of spectral problem for the equation of the
second order is proved.
\end{abstract}
\newpage\

\section{Introduction}

  It is well known that spectral problems for ordinary differential equations
(O.D.E) begin with G.D.Birkhoff's work [1], where the reqular boundary
conditions were determined, it was established an expansion formula of
function in series on eigen and joint elements (E.J.E) of relative
operators, when the coefficionts of equation and boundary conditions don't
depend on the complex parameter $\lambda$.

The following more general result belongs to Ya.D.Tamarkin [2], where a
class of regular and strong regular spectral problems was studied,the
theorems on expansion of functions in series on their root elements (i.e. on
E.J.E) are proved, when the coefficients of equation and boundary conditions
depend on the complex parameter $\lambda$.

In [4-6] for a spectral problem considered in [2] , but for multipoint with
discontinuous coofficients, the notion of regularity is given by
M.L.Rasulov, distinct from [2] multiple expansion formula is obtained shat
it's extremely important with point of view of solving of corresponding
mixed problems for partial differential equations.

In M.V.Keldyshs article [7] it was introducted the theorem on multiple completeness of a
system E.J.E. of a spectral problem which covers some non-regular cases.
But in A.A.Shkalikov sarticle [8] the wide subclass of
non-regular problems (called almost regular) for which the system of root
elements has a block-basisness property in determined spaces, is chosen.
However, in [8] the almost regularity conditions of a spectral problems are
connected with geometry of dominating addends of asymptotical expansion of
characteristic determinant of Green's function that doesn't allows in the
form of the problem to clearify, if it's almost regular and if it's , then
of which order?

The laborionus and cumbersome calculations of coefficients of asymptotical
expansion of characteristic determinant in the case of variabls of
coefficients of equation, be practically impossible. Therefore, this
question was open.

In this paper for a spectral problem of second order equation we obtained a
necessary and sufficient condition of almost regularity of $m$ (arbitrary)
order, distinct from [8] , allowiny immediately and elementarily to answer
to the question of almost regularity of a spectral problem and determine
order of almost regularity. It's proved that for equations with variable
coefficients the almost regular problems of any order exist.

\section{Statement of a problem and some preliminary remarks}

  Consider the spectral problem

\begin{equation}
\label{(1)}l (\frac d{dx},\lambda) y\equiv y^{\prime \prime
}+q(x) y=\lambda ^2y,\,\,\,\,\,\,\,\,\,\,0<x<1
\end{equation}

\begin{equation}
\label{(2)}U_i(y) \equiv \sum_{j=0}^1 \alpha_{ij}y^{(j)}(0)+ \beta _{ij}y^{(j)}(1) =0
\end{equation}
where $q\left(x\right)$ is a complex-valued function, $\alpha _{ij}$,
$\beta_{ij}$ are complex numbers such that \\
${\alpha _{11}\alpha_{12}\beta _{11}\beta _{12}}{\alpha _{21}\alpha _{22}\beta
_{21}\beta _{22}}=2$, $\lambda$ is a spectral parameter.

It's known [3] that if $\alpha _{11}\beta _{21}-\beta _{11}\alpha _{21}\neq
0 $, then the conditions (2) are regular by Birkhoff. If $\alpha _{11}\beta
_{21}-\beta _{11}\alpha _{21}=0$, then without loss of generality we can
assume that $\alpha _{21}=\beta _{21}=0$. Where in the last case, the
boundary conditions (2) will be regular by Birkhoff only on fulfilment of
one of the conditions $\alpha _{11}\beta _{20}+\beta _{11}\alpha _{20}\neq 0$%
, $\alpha _{11}=\beta _{11}=0$, $\alpha _{10}\beta _{20}-\beta _{10}\alpha
_{20}\neq 0$.

This, among the various boundary conditions of the form (2) which are only
non-regular by Birkhoff that they may be led to the form

\begin{equation}
\label{3}
\begin{array}{c}
$$
U_1(y)=\alpha _{11}y^{\prime}(0) +\alpha _{10}y(0)+\beta _{11}y^{\prime}(1) +\beta _{10}y(1) =0, \\

U_2(y)=\alpha _{20}y(0) +\beta _{20}y(1)=0
$$

\end{array}
\end{equation}
and for whose coefficients

\begin{equation}
\label{4}\alpha _{11}\beta _{20}+\beta _{11}\alpha _{20}=0, |\alpha_{11}| +|\beta _{11}| >0, |\alpha _{20}|+|\beta _{20}| >0
\end{equation}
is satisfied.

Thus, we'll consider the problem (1), (3) on fulfilment of the condition (4).

\section{Connection between periodic properties of the potential $q(x)$ of
equation and almost regularity of any order of a spectral problem}

Here under definite conditions on smoothness of the function $q(x)$ , we
obtain a formula allowing us to calculate the coefficients of any member of
asymptotic expansion on power $\lambda $ (when $\left| \lambda \right|
\rightarrow \infty $) of fundamental system of particular solutions
(f.s.p.s) of the equation (1).

Let, $q(x)\in C^m[0,1]$ , by the method of reduction to the systems of
integral equations [3] , it's shown that for sufficiently large $R>0$ at
each of the domains $C_R^{+}=\{{\lambda^{\prime} Re\lambda >0, |\lambda| >R}\},
C_R^{-}=\{{\lambda^{\prime} Re\lambda<0, |\lambda|>R}\}$ of the equation has
f.s.p.s, allowing the asymptotical representations
\begin{equation}
\label{5}\frac{d^\nu y_i(x,\lambda)}{dx^\nu}=\lambda^\nu\exp[(-1)^i \lambda x]
[\sum_{s=0}^m \lambda^{-s}g_{i\nu}^{(s)}(x)+\eta _{i\nu }(x,\lambda)],\,\,
(i=1,2;\nu=0,1)
\end{equation}
where the functions $\eta _{i\nu }(x,\lambda)$ are continuous at $x\in
[0,1] $, analitical by $\lambda \in C_R^{\pm }$ and satisfy the inequality
\begin{equation}
\label{6}|\eta _{i\nu }(x,\lambda )|\leq M|\lambda|^{-m-1}
\end{equation}
and $g_{i\nu }^{(S)}(x)$ are determined by the recursion relations%
$$
g_{i\nu }^{(0)}(x)=(-1)^\nu ,\,\,g_{i\nu}^{(1)}(x)=(-\frac 12)^{i(\nu +1)-1}%
\int_{0}^x q(\xi)d\xi ,\,\,(i=1,2;\nu =0,1)
$$
$$
g_{i\nu }^{(s)}(x)=(-\frac 12)^{i(\nu +1)-1}
\int_{0}^x q(\xi)g_{i0}^{(s-1)}(\xi)d\xi +%
\sum_{j=0}^{s-2} (-\frac 12)^{(i-1)(\nu+s-j)}\frac{d^{s-j-2}}{dx^{s-j-2}} \\
$$
$$
\times [q(x)g_{i0}^{(j)}(x)],
$$
\begin{equation}
\label{7}(i=1,2;\nu =0,1;s=\overline{2,m)}
\end{equation}

Further, using the formula (5) and estimation (6) for characteristic
determinant of Green's problem (1),(3) we obtain the expression
\begin{equation}
\label{8}\triangle(\lambda)=\delta _{-1}(\lambda )e^{-\lambda }+\delta
_0(\lambda )+\delta _1(\lambda )e^\lambda ,
\end{equation}
\begin{equation}
\label{9}\delta _k(\lambda )=\sum_{i=0}^m \lambda^{1-i}\delta_k^{(1-i)}+
O(\frac 1{\lambda^m}),\,\,\,(k=-1,0,1)
\end{equation}
where $\delta _k^{(1-i)}\,\,\,\,(k=-1,0,1;\,i=0,1,...m)$ are constants
given by the formulas
$$
\delta _{-1}^{(1-i)}=\sum_{j=0}^i [\alpha_{20}\beta _{11}g_{20}^{(j)}(0)g_{11}^{(i-j)}(1)-
\beta _{20}\alpha_{11}g_{21}^{(j)}(0)g_{10}^{(i-j)}(1)] +
$$
$$
+(\alpha _{20}\beta _{10}-\beta _{20}\alpha _{10}) \sum_{j=0}^{i-1}g_{20}^{(j)}(0)g_{10}^{(i-j-1)}(1),
$$
$$
\delta _0^{(1-i)}=\sum_{j=0}^i [\alpha _{20}\alpha_{11}(g_{20}^{(j)}(0)g_{11}^{(i-j)}(0)-g_{10}^{(j)}(0)g_{21}^{(i-j)}(0))+
$$
$$
+\beta _{20}\beta_{11}(g_{20}^{(j)}(1)g_{11}^{(i-j)}(1)-g_{10}^{(j)}(1)g_{21}^{(i-j)}(1))],
$$
$$
\delta _1^{(1-i)}=\sum_{j=0}^i [\beta_{20}\alpha _{11}g_{11}^{(j)}(0)g_{20}^{(i-j)}(1)-\alpha _{20}\beta
_{11}g_{10}^{(j)}(0)g_{21}^{(i-j)}(1)]+
$$

\begin{equation}
\label{10}+(\alpha _{10}\beta _{20}-\beta _{10}\alpha _{20})\sum_{j=0}^{i-1}g_{10}^{(j)}(0)g_{20}^{(i-j-1)}(1),
\end{equation}

$$
(i=0,1,2...,m)
$$

From the formula ((10) and (7)) we find:
\begin{equation}
\label{11}\delta _{-1}^{(1-i)}=(-1)^i\delta _1^{(1-i)},\,\,\,(i=1,1,2,...m)
\end{equation}

From the formulas (11), (8), (9) according to A.A.Shkalikov's definition [8]
we have
Definition 1. Let $q(x)\in C^m[0,1]$, the boundary forms $U_i(i=1,2)$
have the form (3) and $|\alpha _{11}| + |\beta_{11}| >0,\,\,
|\alpha _{20}| +|\beta _{20}| >0$. Then the spectral problem (1), (2) is called
almost regular of the order $m\geq 0$, if
\begin{equation}
\label{12}\delta _{-1}^{(1)}=\delta _{-1}^{(0)}=...=\delta
_{-1}^{(2-m)}=0,\,\,\delta _{-1}^{(2-m)}\neq 0
\end{equation}

Remark 1. The almost regular problem of zero order ($m=0,\delta
_{-1}^{(1)}=0 $) is regular by Tamarkin-Rasulov and corresponding to it the
boundary conditions (2) are regular by Birkhoff. This follows from the expansion

\begin{equation}
\label{13}\delta _{-1}^{(1)}=-(\alpha _{20}\beta _{11}+\beta _{20}\alpha_{11}),
\end{equation}
that is obtained by the substitution of (7) in (11). Substituting (7) in
(11) when $i=1$ we obtain
\begin{equation}
\label{14}\delta _{-1}^{(0)}=-\frac 12(\alpha _{20}\beta _{11}+\beta_{20}\alpha _{11})
\int_{0}^1 q(\xi)d\xi -(\alpha_{10}\beta _{20}-\beta _{10}\alpha _{20})
\end{equation}
from (13), (14) and definition 1 it's obvious that the problem (1), (2) will
be almost regular of the first order, iff
\begin{equation}
\label{15}\alpha _{20}\beta _{11}+\beta _{20}\alpha _{11}=0,\,\,\alpha
_{10}\beta _{20}-\beta _{10}\alpha _{20}\neq 0
\end{equation}
at the same time for almost regularity of this problem, of the order $m\geq
2 $, it's necessary that
\begin{equation}
\label{16}\alpha _{20}\beta _{11}+\beta _{20}\alpha _{11}=0,\,\,\alpha
_{10}\beta _{20}-\beta _{10}\alpha _{20}=0
\end{equation}

But on fulfilment of (16), the expression for the number $\delta
_{-1}^{(1-i)}\,\,($ $i\geq 2)\,\,$is simplified and takes the form
\begin{equation}
\label{17}\delta _{-1}^{(1-i)}=\alpha _{20}\beta _{11}\sum_{j=0}^i (-1)^j
[g_{10}^{(j)}(0)g_{11}^{(i-j)}(1)-g_{11}^{(j)}(0)g_{10}^{(i-j)}(1)]
\end{equation}
$$
(i=2,3,...),
$$
where it must be $\alpha _{20}\beta _{11}\neq 0$. Since it's easy to see
that on non-fulfilment of this inequality, the problem (1), (2) becomes to
the cauchy problem and isn't normal in sense of [8].
As it's obvious from (17) the order of almost regularity higher than first,
doesn't depend on coefficients of the boundary conditions (2) and may be
connected only with the coefficient $q(x)$ of the equation (1).
From (17) after simple transformations we obtain
$$
\delta _{-1}^{(-1)}=\frac 12\alpha _{20}\beta _{11}[q(1)-q(0)],
$$
$$
\delta _{-1}^{(-2)}=\frac 14\alpha _{20}\beta _{11}\{[q^{\prime
}(1)+q^{\prime }(0)]+[q(1)-q(0)]\int_{0}^1 q(\xi)d\xi\},
$$
$$
\delta _{-1}^{(-3)}=\frac 18\alpha _{20}\beta _{11}\{[q^{\prime \prime
}(1)-q^{\prime \prime }(0)][q^{\prime }(1)+q^{\prime }(0)]\int_{0}^1 q(\xi)d\xi +
$$
\begin{equation}
\label{18}+2[q(1)-q(0)]\int_{0}^1 q(\xi)\cdot g_{10}^{\prime\prime}(\xi)d\xi +2[q^2(1)-q^2(0)],
\end{equation}
and so on. The regularity which we can note from these formulas, is
generalized in the the following proposition: \\
Lemma 1. Let (16) be satisfied, $\alpha _{20}\beta _{11}\neq 0$ and $\delta
_{-1}^{(-1)}=\delta _{-1}^{(-2)}=...=\delta _{-1}^{(-k)}=0$. Then in order
to be $\delta _{-1}^{(-k-1)}=0$, it's necessary and sufficient that
\begin{equation}
\label{19}q^{(k)}(0)=(-1)^kq^{(k)}(1),\,\,\,\,\,\,\,\,(k\geq 0)
\end{equation}
For the proof of this statement we must establish the following lemma, which
has an independent value in sense of simplification of the recursion
relations (7). \\
Lemma 2. For any natural $s$ for the function $g_{10}^{(s)}(x)$ the
following representation is valid.%
$$
g_{10}^{(s)}(x)=2^{-s}\sum_{\nu=1}^{s_0}[{k_1+...+k_\nu =s+1-2\nu }
{\sum}\alpha _{s,k_1,...k_\nu}^{(\nu)}\cdot q^{(k_1)}(x)......q^{(k_\nu)}(x)+
$$
$$
+{k_1+...+k_\nu =s+1-2\nu}{\sum}\alpha_{s,k_1,...k_{\nu -1}}^{(\nu )}
\cdot q^{(k_1)}(x)q^{(k_{\nu -1})}(x)\cdot\cdot \\
$$
\begin{equation}
\label{20}q^{(s-2\nu -k_1-...-k_{\nu -1})}(x)]
\end{equation}
where $s_0=[\frac{s+1}2]$ is real part of $\frac{s+1}2$,
\begin{equation}
\label{21}q_i(x)=2^i \int_{0}^x q(\xi)g_{10}^{(i)}(\xi )d\xi ,\,\,\,\,\,(i=0,1...)
\end{equation}
and the natural numbers $\alpha _s^{(\nu)}...$, are determined by the formulas
\begin{equation}
\label{22}\alpha _{s,k_1,...,k_\nu}^{(\nu)}= \sum_{j=k_\nu+1}^{s-2(\nu-1)-k_1-...k_{\nu-1}}
C_{s-2(\nu-1)-k_1-...-k_{\nu -2}-j}^{k_\nu -1} \alpha _{s,k_1,...k_{\nu -2},j},\,
\end{equation}
$$
\alpha _{s,k_1}^{(1)}=1
$$

Proof: It's obvious that under the designation of (21) we can represent (7)
(when $\nu =0$) in the form of
\begin{equation}
\label{23}g_{10}^{(s)}(x)=2^{-s}\sum_{i=0}^{s-1} q_i^{(s-1-i)}(x),
\end{equation}
whence we have
\begin{equation}
\label{(24)}q_s(x)=2^sq(x)g_{10}^{(s)}(x)=q(x)\sum_{i=0}^{s-1}q_i^{(s-1-i)}(x).
\end{equation}
From (23) subject to (24) we obtain
$$
g_{10}^{(s)}(x)=2^{-s}\{[q_{s-1}(x)+q_0^{(s-1)}(x)] +\sum_{i_1=0}^{s-1}[q_{i_1}^{\prime}(x)]
^{(s-2-i_1)}\} =
$$
$$
=2^{-s}\{[q_{s-1}(x)+q_0^{(s-1)}(x)] +\sum_{i_1=0}^{s-1}q(x) \sum_{i_2=0}^{i_1-1}
[q_{i_2}^{(i_1-1-i_2)}(x)]^{(s-2-i_1)}\} =
$$
$$
=2^{-s}\{[q_{s-1}(x)+q_0^{(s-1)}(x)] +\sum_{i_1=0}^{s-1} \sum_{i_2=0}^{i_1-1}
\sum_{k_1=0}^{s-i_1-2} C_{s-i_1-2}^{k_1} q^{(k_1)}(x) \times q_{i_2}^{(s-3-i_2-k_1)}(x)\}
$$
$$
=2^{-s}\{[q_{s-1}(x)+q_0^{(s-1)}(x)] +\sum_{k_1=0}^{s-3} \sum_{i_2=0}^{s-3-k_1}
\sum_{i_1=i_2+1}^{s-3-k_1}C_{s-i_1-2}^{k_1}q^{(k_1)}(x)q_{i_2}^{(s-3-i_2-k_1)}(x)\}
$$
denoting
$$
\alpha _{s,s-1}^{(1)}=\alpha _{s,0}^{(1)}=1,\,\,\alpha _{s,k_1,i_2}^{(2)}=%
\sum_{i_1=i_2+1}^{s-2-k_1}C_{s-i_1-2}^{k_1}\alpha_{s,i_1}^{(1)}=
\sum_{i_1=i_2+1}^{s-2-k_1}C_{s-i_1-2}^{k_1}=C_{s-i_1-2}^{k_1+1}
$$
from the last formula we have
$$
g_{10}^{(s)}(x)=2^{-s}\{[\alpha _{s,s-1}^{(1)}q_{s-1}(x)+\alpha
_{s,0}^{(1)}q_0^{(s-1)}(x)] +\sum_{k-1+i_2 \leq s-3}\alpha_{s,k_1,i_2}^{(2)}q^{(k_1)}(x) \\
$$
$$
\cdot q_{i_2}^{(s-3-i_2-k_1)}(x)\} = 2^{-s}\{[\alpha _{s,s-1}^{(1)}q_{s-1}(x)+\alpha_{s,0}^{(1)}q_0^{(s-1)}(x)] + \\
$$
$$
[\sum_{k_1+i_2=s-3}\alpha _{s,k_1,i_2}^{(2)}q^{(k_1)}(x)q_{i_2}(x)+ \sum_{k_1=0}^{s-4}\alpha _{s,k_1,0}^{(2)}\cdot q^{(k_1)}(x) \\
$$
\begin{equation}
\label{25}\times q_0^{(s-3-k_1)}(x)] \sum_{k_1+i_2 \leq S-4}\alpha
_{s,k_1,i_2}^{(2)}q^{(k_1)}(x)q_{i_2}^{(s-3-i_2-k_1)}(x)\}
\end{equation}
using the formulas (23),(24) we transform the last addend in the right hand
side of (25)
$$
\sum_{k_1+i_2 \leq S-4}\alpha_{s,k_1,i_2}^{(2)}q^{(k_1)}(x)q_{i_2}^{(s-3-i_2-k_1)}(x)=
$$
$$
=\sum_{k_1+i_2 \leq S-4}\alpha_{s,k_1,i_2}^{(2)}q^{(k_1)}(x)[q(x)
\sum_{i_3=0}^{i_2-1}q_{i_3}^{(i_2-1-i_3)}(x)]^{(s-4-k_1-i_2)}=
$$
$$
=\sum_{k_1=0}^{s-5} \sum_{i_2=1}^{s-4-k_1} \sum_{i_3=0}^{i_2-1}
\sum_{k_2=0}^{s-4-k_1-i_2}C_{s-4-k_1-i_2}^{k_2}\alpha
_{s,k_1,i_2}^{(2)}q^{(k_1)}(x)q^{(k_2)}(x)\times
$$
$$
\times q_{i_3}^{(s-5-k_1-k_2-i_3)}(x)=\sum_{k_1=0}^{s-5}\,\,\sum_{k_2=0}^{s-5-k_1}\,\,
\sum_{i_3=0}^{s-5-k_1-k_2}\,\,\sum_{i_2=i_3+1}^{s-4-k_1-k_2}C_{s-4-k_1-i_2}^{k_2}\times
$$
$$
\times \alpha_{s,k_1,i_2}^{(2)}q^{(k_1)}(x)q^{(k_2)}(x)q_{i_3}^{(s-5-k_1-k_2-i_3)}(x)=
$$
\begin{equation}
\label{26}=\sum_{k_1+k_2+i_3 \leq S-5}\alpha_{s,k_1k_2,i_3}^{(3)}q^{(k_1)}(x)q^{(k_2)}(x)q_{i_3}^{(s-5-k_1-k_2-i_3)}(x)
\end{equation}
where
$$
\alpha _{s,k_1,k_2,i_3}^{(3)}=\sum_{i_2=i_3+1}^{s-4-k_1-k_2}C_{s-4-k_1-i_2}^{k_2}\alpha _{s,k_1,i_2}^{(2)}.
$$
substituting (26) in (25) we obtain
$$
g_{10}^{(s)}(x)=2^{-s}\{[\alpha_{s,s-1}^{(1)}q_{s-1}(x)+\alpha
_{s,0}^{(1)}q_0^{(s-1)}(x)] +
$$
$$
+[\sum_{k_1+i_2=s-3}\alpha_{s,k_1,i_2}^{(2)}q^{(k_1)}(x)q_{i_2}(x)+\sum_{k_1\leq s-4}\alpha _{s,k_1,0}^{(2)}q^{(k_1)}(x)q_0^{(s-3-k_1)}(x)] +
$$
$$
+[\sum_{k_1+k_2+i_3=s-5}\alpha_{s,k_1,k_2,i_3}^{(s)}q^{(k_1)}(x)q^{(k_2)}(x)q_{i_3}(x)+
$$
$$
+\sum_{k_1+k_2\leq s-6}\alpha_{s,k_1,k_2,0}^{(3)}q^{(k_1)}(x)q^{(k_2)}(x)q_0^{(s-5-k_1-k_2)}(x)] +
$$
$$
+\sum_{k_1+k_2+i_3\leq s-6}\alpha_{s,k_1,k_2,i_3}^{(3)}q^{(k_1)}(x)q^{(k_2)}(x)q_{i_3}^{(s-5-k_1-k_2-i_3)}(x)\}
$$
By repeating the above mentioned transformation once more $S_0-3$ times, it's
evident that we obtain the formulas (20),(22) that it's easily confirmed
also by induction.
Remark 2. The natural numbers $\alpha _{s,k_1,...,k_\nu}^{(\nu)}$
appearing in the formula (20), have the propety
\begin{equation}
\label{27}\alpha _{s,k_1,...,k_\nu}^{(\nu)}=\alpha _{s-k_\nu
,k_1,...,k_{\nu -1},0}^{(\nu)}
\end{equation}
which will be usefull to prove lemma 1.
Proof of lemma 1. We use the mathematical induction method. When $k=0$ and
$n=1$ the validity of statement of lemma follows from the first two
equalities of the formula (18). Let for some $i\geq 4$ it be valid for all
$k\leq i-3$. We proof the validity of statement of lemma when $k=i-2$ , i.e.
let the conditions (16) be satisfied, $\alpha _{20}\cdot \beta _{11}\neq 0$
and
\begin{equation}
\label{28}\delta _{-1}^{(-1)}=...=\delta _{-1}^{(2-i)}
\end{equation}
and we establish that $\delta _{-1}^{(1-i)}=0$, iff the following equality
holds
\begin{equation}
\label{29}q^{(i-2)}(1)=(-1)^{i-2}q^{(i-2)}(0),
\end{equation}
from the fulfilment of (28) by virtue of our supposition it follows that
\begin{equation}
\label{30}q^{(k)}(1)=(-1)^kq^{(k)}(0),\,\,(k=0,1,...,i-3).
\end{equation}
It's lasy to see that by using the designation (21) we can represent in the
form of
$$
(\alpha_{20}\beta_{11})^{-1}\delta _{-1}^{(1-i)}=2[g_{10}^{(i)}(1)+(-1)^{i+1}g_{10}^{(i)}(0)] +
$$
\begin{equation}
\label{31}+\sum_{j=0}^{i-1}(-1)^{j+1}2^{1+j-i}g_{10}^{(j)}(0)q_{i-j-1}(1),
\end{equation}
substituting (20) in (31) we find
$$
2^{i-1}(\alpha _{20}\beta _{11})^{-1}\delta _{-1}^{(1-i)}=
\sum_{\nu=1}^{i_0}\{\sum_{k_1+...k_\nu=i+1-2\nu }\alpha_{i,k_1,...k_{\nu -1}}^{(\nu)}\times
$$
$$
\times [q^{(k_1)}(1)...q^{(k_{\nu -1})}(1)q_{k\nu}(1)+(-1)^{i+1}q^{(k_1)}(0)...q^{(k_{\nu -1})}(0)\times q_{k\nu}(0)]+
$$
$$
+\sum_{k_1+...+k_\nu \leq 1-2\nu}\alpha_{i,k_1,...k_{\nu -1,}0}^{(\nu)}[q^{(k_1)}(1)...q^{(k_{\nu-1})}(1)
q^{(1-2\nu -k_1-...-k_{\nu -1})}(1)+
$$
$$
q^{(k_1)}(0)...q^{(k_{\nu-1})}(0)q^{(1-2\nu -k_1-...-k_{\nu -1})}(0)]\}-q_{i-1}(1)+
$$
$$
\sum_{j=2}^{i-1}(-1)^{j+1}q_{i-j-1}(1)\sum_{\nu=1}^{i_0}[\sum_{k_1+...+k_\nu}^{j+1-2\nu}\alpha _{j,k_1,...k_\nu }^{(\nu)} \times
$$
$$
q^{(k_1)}(0)...q^{(k_{\nu -1})}(0)\,\,q_{k\nu }(0)+
$$
\begin{equation}
\label{32}+\sum_{k_1+...k_\nu \leq j-2\nu} \alpha_{j,k_1,...k_{\nu -1},0}^{(\nu)}q^{(k_1)}(0)...q^{(k_{\nu
-1})}(0)\,q^{(j-2\nu -k_1-...-k_{\nu -1})}(0)],
\end{equation}
where we denote by $i_0$ and $j_0$, integer parts of the numbers $\frac{i+2}{2}$ and $\frac{j+1}{2}$, \\
respectively. Subject to (30) and (21) $q_{k \nu}(0)=0$
(see (21)) from (32) \\ we have
$$
2^{i-1}(\alpha_{20}\beta _{11})^{-1}\delta _{-1}^{(1-i)}=%
\sum_{\nu =1}^{i_0}\,\,\sum_{k_1+...+k_\nu=i+1-2\nu}\alpha _{i,k_1,...k_{\nu -1}}^{(\nu)}\times
$$
$$
\times q^{(k_1)}(1)...q^{(k_{\nu -1})}(1)q_{k\nu }(1)+\sum_{\nu =2}^{i_0}\,\,\sum_{k_1+...k_\nu =i+1-2\nu}
\alpha_{i,k_1,...k_\nu ,0}^{(\nu )}\times
$$
$$
\times q^{(k_1)}(1)...q^{(k_{\nu -1})}(1)q^{(i-2\nu -k_1-...-k_{\nu
-1})}(1)\times
$$
$$
\times [1+(-1)^{i+1}(-1)^{1-2\nu}] +\alpha _{i,0}^{(1)}[q^{(i-2)}(1)+(-1)^{i+1}q^{(i-2)}(0)] -
$$
$$
-q_{i-1}(1)+\sum_{j=2}^{i-1}(-1)^{j+1}q_{i-j-1}^{(1)}\sum_{\nu =1}^{j_0}\,\,\,
\sum_{k_1+...+k_{\nu-1}\leq j-2\nu}\alpha _{j,k_1,...k_\nu ,0}^{(\nu )}\times
$$
$$
\times q^{(k_1)}(0)...q^{(k_{\nu -1})}(0)q^{(j-2\nu -k_1-...-k_{\nu-1})}(0)=
$$
$$
=\alpha _{i,0}^{(1)}[q^{(i-2)}(1)+(-1)^{i+1}q^{(i-2)}(0)] +
$$
$$
+\sum_{\nu =2}^{i_0}\,\,\,\sum_{k_1+...k_\nu=i+1-2\nu}\alpha _{i,k_1,...k_\nu}^{(\nu)}
q^{(k_1)}(1)...q^{(k_{\nu -1})}(1)q_{k\nu}(1)+
$$
$$
+\sum_{j=2}^{i-1}(-1)^{j+1}q_{i-j-1}^{(1)}\sum_{\nu =1}^{j_0}\,\,\,\sum_{k_1+...k_\nu \leq j-2\nu }
(-1)^{j+2\nu}\alpha _{j,k_1,...k_{\nu -1},0}^{(\nu)}\times
$$
\begin{equation}
\label{33}\times q^{(k_1)}(1)...q^{(k_{\nu -1})}(1)q^{(j-2\nu
-k_1-...-k_{\nu -1})}(1)
\end{equation}
By virtue of the formula (22), $\alpha _{i,0}^{(1)}=0$ and when
$k_1+...+k_\nu = \\ i+1-2\nu$ we have;
$$
\alpha _{i,k_1,...k_\nu }^{(\nu)}=\sum_{j=i+2-2\nu -k_1-...-k_{\nu -1}}^{i-2(\nu -1)-k_1-...-k_{\nu -1}}C_{i-2(k-1)-k_1-...-k_{\nu -1}-j}^{k_{\nu -1}}\times
$$
$$
\times \alpha _{i,k_1,...k_{\nu -2},j}^{(\nu -1)}=C_{k_{\nu -1}}^{k_{\nu
-1}}\alpha _{i,k_1,...k_{\nu -2},i-2(\nu -1)-k_1-...-k_{\nu -1}}=
$$
$$
=\alpha _{i,k_1,...k_{\nu -2},i-2(\nu -1)-k_1-...-k_{\nu -1}}^{(\nu -1)},
$$
substituting these values in the right Land side of (33) we obtain%
$$
2^{i-1}(\alpha _{20}\beta _{11})^{-1}\delta_{-1}^{(1-i)}=[q^{(i-2)}(1)+(-1)^{i+1}q^{(i-2)}(0)] +
$$
$$
+\sum_{\nu =2}^{i_0}\,\,\,\sum_{k_1+...k_\nu=i+1-2\nu}\alpha _{i,k_1,...k_{\nu -2},i-2(\nu -1)-k_1-...-k_{\nu
-1}}^{(\nu -1)}\times
$$
\begin{equation}
\label{34}\times q^{(k_1)}(1)...q^{(k_{\nu -1})}(1)q_{k\nu }(1)-\sum_{j=2}^{i-1}q_{i-j-1}(1)\times
\end{equation}
$$
\times \sum_{\nu =1}^{j_0}\,\,\,\sum_{k_1+...+k_{\nu -1}\leq j-2\nu}\alpha _{j,k_1,...k_{\nu -1},0}^{(\nu
)}q^{(k_1)}(1)...q^{(k_{\nu -1})}(1)\times
$$
$$
\times q^{(j-2\nu -k_1-...-k_{\nu -1})}(1)
$$
we prove that
$$
\sum_{\nu =2}^{i_0}\,\sum_{k_1+...k_\nu=i+1-2\nu } \alpha _{i,k_1,...k_{\nu -2},i-2(\nu -1)-k_1-...-k_{\nu
-1}}^{(\nu -1)}\times
$$
$$
\times q^{(k_1)}(1)...q^{(k_{\nu -1})}(1)q_{k\nu }(1)=
$$
\begin{equation}
\label{35}=\sum_{j=2}^{i-1}q_{i-j-1}(1)\sum_{\nu =1}^{j_0}\,\sum_{l_1+..+l_{\nu -1}}^{j-2\nu}
\alpha _{j,l_1,...l_{\nu -1},0}^{(\nu)}\times
\end{equation}
$$
\times q^{(l_1)}(1)...q^{(l_{\nu -1})}(1)q^{(j-2\nu -l_1-...-l_{\nu -1})}(1)
$$
It's easy to see that the orders of derivatives and the indices $q(1)$
in left and right hand sides of the equality (35) takes the same values.
Therefore it's sufficient to show the equality of correspanding
coefficients. Let's fix some numbers $\nu =m$ and $k_1=p_1,...,\,\,k_\mu
=p_\mu$ such that $2\leq \mu \leq i_0,\,\,p_1+...+p_\mu =i+1-2\mu$. In the
left hand side of (35) the coefficient of production $q^{(p_1)}(1)...q^{(p_{%
\mu -1})}(1)q_{p_\mu }(1)$ will be the number
\begin{equation}
\label{36}\alpha _{i,p_1,...p_{\mu -2},i-2(\mu -1)-p_1-...-p_{\mu -1}}^{(\mu-1)},
\end{equation}
and in the right hand side, such production arises when \thinspace $%
j=i-p_m-1,\,\,\nu =\,=\mu -1,\,\,l_1=p_1,\,\,...,\,l_{\mu -2}=p_{\mu
-2},\,\,\,j-2(\mu -1)-l_{\mu -1}-...-l_{\mu -2}=p_{\mu -1}\,$(i.e. $p_{\mu
-1}=i-p_\mu -1-2(\mu -1)-p_1-...-p_{\mu -2}$). In addition, the coefficient
of this production will be the number
\begin{equation}
\label{37}\alpha _{i-p_\mu -1,p_1,...,p_{\mu -2},0}=\alpha _{2(\mu
-1)+p_1+...+p_{\mu -1},p_1,...,p_{\mu -2},0}^{(\mu -1)}
\end{equation}
The last equality is valid in connection with fact that $p_1+...+p_{\mu
-1}=i+1-2\mu $ and the right hand side of (35) are equal to the number (36)
by virtue of Remark 1. Then from (34) we have
\begin{equation}
\label{38}2^{i-1}(\alpha _{20}\beta _{11})^{-1}\delta_{-1}^{(1-i)}=q^{(i-2)}(1)+(-1)^{i+1}q^{(i-2)}(0)
\end{equation}
from (38) it follaws the validity of statement that $\delta _{-1}^{(1-i)}=0$
iff (29) satisfied Lemma 1 is proved. From lemma1, definition 1 and the
formulas (13), (14) the validity of the following basic statement
immediately follows.

Theorem. Let $q(x)\in C^m[0,1]$, the boundary forms $U_i\,\,(i=1,2)$ have
the form (3) and $|\alpha _{11}| +|\beta _{11}|>0,\,\,|\alpha _{20}| +|\beta _{20}| >0$.
Then for almost regularity of the order $m\geq 0$ of the spectral problem (1), (2),
it's necessery and sufficient that:
$\alpha _{11}\beta _{20}+\beta _{11}\alpha _{20}\neq 0,$
when $m=0$ (regularity)
$\alpha _{11}\beta _{20}+\beta _{11}\alpha _{20}=0,\,\,\,\alpha _{10}\beta
_{20}-\beta _{10}\alpha _{20}\neq 0,\,\,\,\,\,\,\,\,\,\,$when $m=1;$
$\alpha _{11}\beta _{20}+\beta _{11}\alpha _{20}=0,\,\,\,\alpha _{10}\beta
_{20}-\beta _{10}\alpha _{20}=0,\,\,\,\,\alpha _{11}\beta _{20}\neq 0,$
$q^{(i)}(0)=(-1)^iq^{(i)}(1),\,q^{(m-2)}(0)=(-1)^{m-2}q^{(m-2)}(1),\,\,\,\,m\geq 2$.
\newpage\


\begin{thebibliography}{99}
\bibitem{birkhoff}  Birkhoff G.D. Boundary value and expansion problems of
ordinary linear differential equations. trans. Amer. Math. soc., 1908, 9,
p.373-395.
\bibitem{tamarkin}  Tamarkin Ya. D. On some general problem of the theory of
ordinary differential equations and on expansion of orbitrary functions in
series. Petrograd, 1917, 204p.
\bibitem{naymark}  Naymark M.A. Linear differential operators. M.,
''Nauka'', 1969, p.
\bibitem{rasulov}  Rasulov M.L. Contour integral methods. M.''Nauka'', 1964,
462p.
\bibitem{rasulov}  Rasulov M.L. Application of residue method. Baku,
''Elm'', 1989, 328p.
\bibitem{rasulov}  Rasulov M.L., Mamedov Yu. A. On the residue method of
solutions of mixed problem for a class of hyperbolic systems. //DAN. SSSR,
1988, v.300, N.6, p.1321-1324.
\bibitem{keldysh}  Keldysh M.V. On cigen values and ligen functions of some
classes of non-seef-adjoint equations. //DAN SSSR, 1951, v.77, N.1, p.11-14.
\bibitem{shralikov}  Shkalikov A.A. Boundary value problems for ordinary
differential equations with a parameter in boundary conditions. //Proceedigs
of I.G.Petrovsky's seminar, M., 1983, issue 9, p.190-229.
\bibitem{mamedov}  Mamedov Yu.A. On spectral problems for a system of
ordinary differential equation not being normal. //DAN SSSR, 1989, v.306,
N.4, p.540-544.
\bibitem{mamedov}  Mamedov Yu.A., Ahmedov Kh. I. On regilarity conditions
and asymptotic of eejen values of a spectral problem. //Collection of papers
of the first Republican conference on mechanics and mathematics declucated
to the 50-th anniversary of \thinspace Academy Science of Azerbaijan (Baku,
5-15June, 1995) Part 1, Mechanics, p.106-109
\end{thebibliography}
\end{document}